\documentclass[twocolumn,trackchanges]{aastex6}
\RequirePackage{lineno}
\usepackage{amsmath}
\usepackage{amssymb}
\usepackage{amsthm}
\usepackage{natbib,aasdefs,url,bm}
\usepackage{array}
\usepackage{float}
\usepackage{graphicx}
\usepackage{subfigure}
\usepackage{color}
\usepackage{xspace}

\newcounter{ichi}
\newcounter{ni}
\newcounter{san}
\newcounter{yon}
\def\be{\begin{equation}}
\def\ee{\end{equation}}
\def\ba{\begin{eqnarray}}
\def\ea{\end{eqnarray}}

\newcommand{\icnu}{IceCube-191001A\xspace}
\newcommand{\tde}{AT2019dsg\xspace}

\def\aap{\ {A\&A}\ }

\def\apjl{\ {ApJL}\ }
\def\apjs{\ {ApJS}\ }

\shorttitle{High-Energy Multi-Messenger Emission from TDEs}
\shortauthors{Murase et al.}

\begin{document}

\title{High-Energy Neutrino and Gamma-Ray Emission from  Tidal Disruption Events}

\author{Kohta Murase\altaffilmark{1,2,3,4}, Shigeo S. Kimura \altaffilmark{5,6}, B. Theodore Zhang\altaffilmark{1,2,3}, Foteini Oikonomou\altaffilmark{7}, Maria Petropoulou\altaffilmark{8}}
\altaffiltext{1}{Department of Physics, Pennsylvania State University, University Park, PA 16802, USA}
\altaffiltext{2}{Department of Astronomy \& Astrophysics, Pennsylvania State University, University Park, PA 16802, USA}
\altaffiltext{3}{Center for Multimessenger Astrophysics, Institute for Gravitation and the Cosmos, Pennsylvania State University, University Park, PA 16802, USA}
\altaffiltext{4}{Center for Gravitational Physics, Yukawa Institute for Theoretical Physics, Kyoto University, Kyoto, Kyoto 606-8502, Japan}
\altaffiltext{5}{Frontier Research Institute for Interdisciplinary Sciences, Tohoku University, Sendai 980-8578, Japan}
\altaffiltext{6}{Astronomical Institute, Tohoku University, Sendai 980-8578, Japan}
\altaffiltext{7}{Institutt for fysikk, NTNU, Trondheim, Norway}
\altaffiltext{8}{Department of Astrophysical Sciences, Princeton University, Princeton, NJ 08544, USA}


\begin{abstract}
Tidal disruption events (TDE) have been considered as cosmic-ray and neutrino sources for a decade. 
We suggest two classes of new scenarios for high-energy multi-messenger emission from TDEs that do not have to harbor powerful jets. 
First, we investigate high-energy neutrino and gamma-ray production in the core region of a supermassive black hole. In particular, we show that $\sim1-100$~TeV neutrinos and MeV gamma-rays can efficiently be produced in hot coronae around an accretion disk. We also study the consequences of particle acceleration in radiatively inefficient accretion flows (RIAFs).  
Second, we consider possible cosmic-ray acceleration by sub-relativistic disk-driven winds or interactions between tidal streams, and show that subsequent hadronuclear and photohadronic interactions inside the TDE debris lead to GeV-PeV neutrinos and sub-GeV cascade gamma-rays. 
We demonstrate that these models should be accompanied by soft gamma-rays or hard X-rays as well as optical/UV emission, which can be used for future observational tests. 
Although this work aims to present models of non-jetted high-energy emission, we discuss the implications of the TDE AT2019dsg that might coincide with the high-energy neutrino IceCube-191001A, by considering the corona, RIAF, hidden sub-relativistic wind, and hidden jet models. It is not yet possible to be conclusive about their physical association and the expected number of neutrinos is typically much less than unity. We find that the most optimistic cases of the corona and hidden wind models could be consistent with the observation of IceCube-191001A, whereas jet models are unlikely to explain the multi-messenger observations. 
\end{abstract}


\section{Introduction}
The new era of multi-messenger particle astrophysics has begun with real-time observations of high-energy neutrinos~\citep[see reviews, e.g.,][]{Halzen:2016gng,Ahlers:2017wkk,Meszaros:2019xej,Murase:2019tjj}. Various attempts to discover transient (bursting or flaring) neutrino sources are ongoing, and include not only electromagnetic follow-up observations but also real-time multi-messenger searches using sub-threshold data~\cite[e.g.,][for Astrophysical Multi-messenger Network Observatory (AMON)]{Smith:2012eu,AyalaSolares:2019iiy}.
In particular, the IceCube Neutrino Observatory~\footnote{http://icecube.wisc.edu/} reported the detection of a $\sim 200$ TeV muon neutrino on 2017 September 22. Follow-up observations revealed that the neutrino, IceCube-170922A, was coincident with the long-duration gamma-ray flare of the blazar TXS 0506+056~\citep{Aartsen2018blazar1}, and a neutrino flare was found in the 2014-2015 data by a subsequent analysis~\citep{Aartsen2018blazar2}.  
Although their physical interpretation of the multi-messenger data has been under debate especially for the 2014-2015 neutrino flare~\citep[e.g.,][]{Murase:2018iyl,2019ApJ...881...46R, Rodrigues:2018tku,Zhang:2019htg,Petropoulou:2019zqp}, it provided a new way to diagnose high-energy phenomena caused by supermassive black holes (SMBHs) and constrain high-energy cosmic-ray (CR) acceleration in powerful jets. 

On 2019 October 1, the IceCube Collaboration reported the detection of another $\sim 200$ TeV muon neutrino, \icnu, with $59\%$ probability of astrophysical origin~\citep{gcn25913}. Although the localization uncertainty of this event is not great ($\sim25.53~{\rm deg}^2$), follow-up observations with the Zwicky Transient Facility~\citep[ZTF,][]{2019PASP..131g8001G} revealed several optical transients~\citep{2019ATel13160....1S} within the error circle of the arrival direction of the neutrino. Among them there was the tidal disruption event (TDE), \tde, observed approximately 150 days post peak. \tde was first detected by the ZTF survey on 2019 April 9~\citep{2019TNSTR.615....1N}, and triggered multi-wavelength follow-up optical, UV, X-ray and radio observations~\citep{2019ATel12777....1P,2019ATel12798....1S,2019ATel12825....1P,2019ATel12960....1P,2019TNSAN..13....1N,vanVelzen:2020cwu}. \tde is one of only a handful of radio-detected TDEs. 

In a joint analysis of the neutrino and electromagnetic observations it was concluded that \tde is the most likely counterpart of \icnu~\citep{Stein:2020xhk}. The chance probability of detecting a high energy neutrino in coincidence with a radio-detected TDE was reported to be $0.5\%$. The presence of a jet in \tde has not been unambiguously established by observations. For example, the X-ray emission of \tde is soft and well described by a blackbody of temperature $\sim 10^{5.9}$~K ($\sim 0.07$~keV), in contrast to the hard non-thermal X-ray emission of jetted TDEs~ \citep[e.g.,][]{2011Natur.476..421B,Bloom:2011xk, 2017ApJ...838..149A}. A time-varying degree of optical polarization in \tde could be associated with a jet, but could also originate from a non-isotropic accretion disk~ \citep{Lee_2020}. 

Theoretically, jetted TDEs were proposed as possible sources of ultrahigh-energy cosmic rays (UHECRs) more than a decade ago~\citep{Farrar:2008ex}, and the associated high-energy neutrino emission was also calculated~\citep{Murase:2008zzc}.
Since then, jetted TDEs have been studied under varying assumptions~\citep{Farrar:2014yla,Zhang:2017hom,AlvesBatista:2017shr,Guepin:2017abw,Biehl:2017hnb}. Neutrino production in TDE environments has been more actively investigated~\citep{Wang:2011ip,Wang:2015mmh,Senno:2016bso,Dai:2016gtz,Lunardini:2016xwi} since the discovery of the first jetted TDE Swift~J1644+57~\citep{2011Natur.476..421B} and the discovery of astrophysical neutrinos by IceCube \citep{PhysRevLett.111.021103,icecubeScience,PhysRevLett.113.101101}.

However, high-energy neutrino emission from TDEs has also been constrained.  
A stacking analysis of IceCube data found no counterparts to previously detected TDEs and concluded that at most $\sim 1\% ~(26\%)$ of IceCube neutrinos may originate in jetted (non-jetted) TDEs~\citep{Stein:2019ivm}. Independently, based on the analyses on point-source emission from Swift~J1644+57 and neutrino multiplets in the IceCube data, \cite{Senno:2016bso} showed that the contribution to the diffuse neutrino flux should be sub-dominant. 

The recent IceCube data in the 10-100~TeV range~\citep{Aartsen:2020aqd} have suggested a population of hidden neutrino sources that are dark in GeV-TeV gamma-rays~\citep{Murase:2015xka,Capanema:2020rjj}. 
TDEs with hidden jets that can be dark in X-rays, have also been considered in the literature~\citep{Wang:2015mmh,Senno:2016bso}. 
Alternatively, \cite{Zhang:2017hom} discussed CR acceleration in sub-relativistic outflows. 
More recently, \cite{Hayasaki:2019kjy} studied possible neutrino emission from radiatively inefficient accretion flows (RIAFs) and magnetically arrested disks (MADs).

In this work we study possible high-energy multi-messenger emission from non-jet regions in TDEs. 
In particular, we investigate ``core'' models, in which high-energy neutrinos and gamma-rays are generated in the vicinity of SMBHs, coronae or RIAFs (Section~\ref{sec:core_models}). We also study a hidden wind model, where particles are accelerated in mildly or sub-relativistic winds inside the TDE debris in Section~\ref{sec:shock_model}. In Section~\ref{sec:AT2019dsg} we discuss the implications of the model predictions for the reported association of \icnu with \tde, including the hidden jet model. 
Finally, we comment on the role of TDEs as possible sources of the diffuse neutrino flux in Section~\ref{sec:diffuse}, and summarize our results in Section~\ref{sec:summary}.  

\begin{figure}[t]
\includegraphics[width=\linewidth]{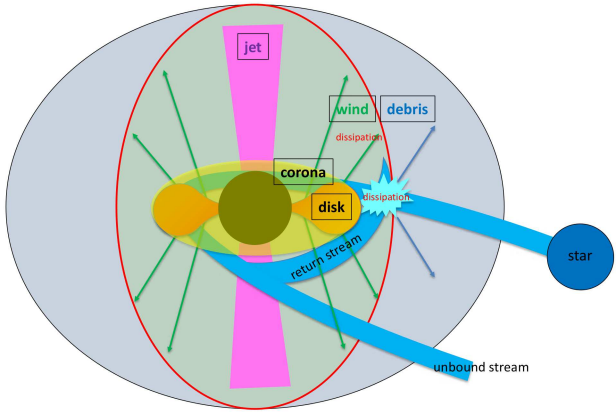}
\caption{Schematic picture of neutrino and gamma-ray production models considered in this work (not to scale). In the core models, the emission region is the corona and disk regions. In the hidden wind model, the emission regions are sub-relativistic outflows that may be driven by an accretion disk or induced by collisions among tidal streams. In the jet model, CR acceleration and neutrino production occur inside relativistic jets. Note that the above scenarios are not mutually exclusive. 
\label{fig:TDE}
}
\vspace{-1.\baselineskip}
\end{figure}

\section{Core Models}
\label{sec:core_models}
TDEs are caused by the disruption of a star, which have been predicted as optical and UV transients~\citep[e.g.,][]{1988Natur.333..523R,Evans:1989qe}. The tidal radius of a black hole with a mass of $M_{\rm BH}\equiv 10^7\,M_{\odot}\, M_{\rm BH,7}$ is estimated to be $R_T\approx{f_{T}}^{1/6}{(M_{\rm BH}/M_*)}^{1/3}R_*\simeq 9.8\times{10}^{12}~{\rm cm}~f_{T,-1.1}^{1/6}M_{\rm BH, 7}^{1/3}M_{*}^{2/3-\xi}$, where $R_*, M_*$ are the radius and mass of the star, $f_{T}\sim0.02-0.3$ is a correction factor related to the shape of the stellar internal density profile~\citep[e.g.,][]{1989IAUS..136..543P,Piran:2015gha}, and $\xi=1-\ln(R_*/R_\odot)/\ln(M_*)$. 

About a half of the disrupted stellar material may fallback and a fraction of the mass would accrete onto a SMBH. 
The fallback time is estimated by the orbital period of the stellar debris on the most bound orbit as
\begin{eqnarray}
t_{\rm fb}\approx2\pi\sqrt{\frac{a_{\rm min}^3}{GM_{\rm BH}}}
\simeq3.2\times{10}^6~{\rm s}\,f_{T,-1.1}^{1/2} M_{\rm BH,7}^{1/2}M_{*}^{\frac{(1-3\xi)}{2}},
\end{eqnarray}
where 
\begin{eqnarray}
a_{\rm min}\approx \frac{R_T^2}{2R_*}\simeq7.0\times{10}^{14}~{\rm cm} f_{T,-1.1}^{1/3} M_{*}^{1/3-\xi}M_{\rm BH,7}^{2/3}
\label{eq:amin}
\end{eqnarray}
is the semi-major axis of the orbit. 
The circularization may start after the most bound debris falls back to the SMBH, which can take a few times longer than $t_{\rm fb}$~\citep[e.g.,][]{Dai:2013hka,Shiokawa:2015iia,Hayasaki:2015pxa,Bonnerot:2015ara}. 
The formation of an accretion disk around the central SMBH has been theoretically expected~\citep[][]{Evans:1989qe,1990ApJ...351...38C, Loeb:1997dv,Strubbe:2009qs} and suggested by observations~\citep[e.g.,][]{Leloudas:2016rmh,vanVelzen:2018dlv,Wevers:2019dxz,Hung:2020jqz}.
The black hole mass accretion rate, which is a function of time, may be expressed as
\begin{equation}
\dot{M}\approx\frac{\eta_{\rm in}\eta_{\rm fb} M_*}{3t_{\rm fb}}{\left(\frac{t}{t_{\rm fb}}\right)}^{-5/3},\,\,\,\,\,\,\,\,\,\,
\label{eq:Mdot-1}
\end{equation}
where $\eta_{\rm fb}$ is the fraction of the stellar debris that falls back, and $\eta_{\rm in}$ is the fraction of inflow mass that ends up forming a disk. These correction factors are rather uncertain, as they depend on the details of circularization and stellar orbits~\citep[e.g.,][]{Shiokawa:2015iia,Guillochon:2015qfa,Svirski:2015fma,Bonnerot:2016cob,Jiang:2016igx,Hayasaki:2018hxv,Lu:2019hwv}, and mass losses due to outflows~\citep[e.g.,][]{Strubbe:2009qs,Miller:2015jda,Metzger:2015sea}.
They can be time dependent, which leads to a deviation from the standard value of $5/3$ for the decay slope of $\dot{M}$. 

For $\eta_{\rm in}\eta_{\rm fb}\gtrsim0.1$, the initial accretion rate is in the super-Eddington regime, i.e., $\dot{M}>L_{\rm Edd}/(\eta_{\rm rad}c^2)$, where $L_{\rm Edd}\approx1.26\times{10}^{45}~M_{\rm BH,7}~{\rm erg}~{\rm s}^{-1}$ is the Eddington luminosity and $\eta_{\rm rad}\sim0.1$ is the radiation efficiency, so accretion through a slim~\citep[e.g.,][]{Strubbe:2009qs,Shen:2013oma} or geometrically thick~\citep[e.g.,][]{Loeb:1997dv,Coughlin:2013lva} disk is expected at early times. 
Once the accretion starts, the accretion flows expand beyond the circularization radius through angular momentum redistribution~\citep{1974MNRAS.168..603L}. 
The viscous time scale of the disk is $t_{\rm vis}\approx\alpha^{-1}\mathcal{H}^{-2}\Omega_K^{-1}\simeq4.4\times10^6~{\rm s}~\alpha_{-1}^{-1}M_{\rm BH,7}\mathcal{H}_{-1}^{-2}(R_d/10 R_S)^{3/2}$, where $\mathcal{H}=H_d/R_d$, $H_d$ is the scale height of the disk, $R_d$ is the disk radius, $\alpha$ is the viscous parameter~\citep{Shakura:1972te}, $\Omega_K$ is the Keplerian angular frequency, and $R_S=2GM_{\rm BH}/c^2$ is the Schwarzschild radius. In the super-Eddington phase, the outflows also affect the disk evolution \citep{Ohsuga:2011jk,Sadowski:2013gua,Jiang:2014tpa}, and the time evolution of the disk radius and mass accretion rate are under debate~\citep[see][where the disk radius much larger than the classical circularization radius has been suggested]{Coughlin:2013lva,Shiokawa:2015iia,Bonnerot:2019yjb}.

When the accretion becomes sub-Eddington, the disk state will change to a standard geometrically-thin/optically-thick disk~\citep{Shakura:1972te}. If the viscous time at circularization radius is longer than the circularization time, the time evolution of the mass accretion rate in the standard disk may be represented by $\dot{M}\approx(L_{\rm Edd}/[\eta_{\rm rad}c^2]){(t/t_{\rm vis})}^{-19/16}$~\citep{1990ApJ...351...38C}, which is applicable to an isolated disk where mass losses or supplies are negligible. 
Note that at earlier times the accretion rate has a shallower index, $-11/14$, due to stalled accretion~\citep{Mummery:2019vgr,Mummery:2019cuh} 
The mass accretion rate is related to the bolometric luminosity as $L_{\rm bol}=\eta_{\rm rad}\dot{M}c^2\simeq1.3\times{10}^{43}\,\eta_{\rm rad,-1}\dot{m}_{-1}M_{\rm BH,7}~{\rm erg}~{\rm s}^{-1}$, where $\dot{m}=\dot{M}c^2/L_{\rm Edd}$ is the normalized mass accretion rate. The transition accretion rate from the super-Eddington to sub-Eddington accretion is given by $\dot{m}=\eta^{-1}_{\rm rad}$, and the viscous time is evaluated at the outer radius of the disk at the state transition. 
Here, we assume that $\dot{M}$ is constant inside the disk, which can be realized if the outflows from the standard disk are negligible, as shown by numerical simulations~\citep[e.g.,][]{Ohsuga:2011jk}. 

If the mass accretion rate decreases below a critical value $\dot{m}_{\rm crit}\approx0.03 \, \alpha_{-1}^2$~\citep{Mahadevan:1997qz}, the accretion state changes into that of a hot accretion flow, or a RIAF.

In the following two subsections, we will explore two core models for high-energy neutrino and gamma-ray production that probe the different accretion regimes described above. 
The neutrino and gamma-ray production sites in the core models are indicated in Figure~\ref{fig:TDE}.

\subsection{Corona model}\label{sec:corona}
By analogy to AGN, we postulate the existence of a hot corona above a slim or standard accretion disk around the central SMBH. The details of long-term disk accretion in the TDE environment are still uncertain~\citep[e.g.,][]{Bonnerot:2015ara}. We estimate plasma quantities and CR properties in coronae using the empirical relations obtained by multi-wavelength observations of AGN~\citep{Murase:2019vdl}. 

Either a slim or standard disk provides copious optical and UV photons, whose spectrum is multi-temperature blackbody emission. In the standard disk, for example, the inner disk temperature is estimated as $T_{\rm disk}\approx0.488\, (3GM_{\rm BH}\dot{M}/8\pi\sigma_{\rm SB}R_{\rm ISCO}^3)^{1/4}$~\citep[e.g.,][]{1981ARA&A..19..137P}, which typically lies in the UV range. 
In the TDE case, the early-time emission may not be directly observed because it can be reprocessed by the surrounding optically thick material~\citep[e.g.,][]{Loeb:1997dv,Strubbe:2009qs,Dai:2018jbr}. But late-time optical and UV emission is often attributed to the disk emission~\citep[e.g.,][]{Leloudas:2016rmh,vanVelzen:2018dlv,Wevers:2019dxz}. 

In a corona, electrons are heated presumably by magnetic dissipation, cooled via the Comptonization of optical/UV disk photons, and efficiently emit hard X-rays. Observationally, the electron temperature in AGN coronae is found to be $kT_e\sim10-100$~keV. 
When the Coulomb relaxation time is longer than the dissipation time scale, one may expect a two-temperature plasma, in which thermal protons have a virial temperature of $kT_p\simeq5.2~{\rm MeV}~{(R/30R_S)}^{-1}$, where $R$ is the coronal radius. The plasma beta, $\beta\equiv\sqrt{8\pi n_p k T_p/B^2}$, is introduced to estimate the magnetic field strength $B$. Here, $n_p$ is the number density of thermal protons. For $\beta\sim0.01-1$, we expect $B\sim0.1-30$~kG.   

For AGNs, there is an empirical relationship between the bolometric luminosity $L_{\rm bol}$ and X-ray luminosity $L_X$ (in the 2-10 keV energy range), which reads $L_X\sim(0.03-0.1)L_{\rm bol}$ for $L_{\rm bol}\sim{10}^{42}-10^{45}~{\rm erg}~{\rm s}^{-1}$~ \citep{Hopkins:2006fq}. The spectral properties of the disk-corona system are often characterized by the Eddington ratio, $\lambda_{\rm Edd}\equiv L_{\rm bol}/L_{\rm Edd}$~ \citep{Ho:2008rf}. The coronal X-ray spectrum becomes softer for larger values of $\lambda_{\rm Edd}$, which is also consistent with the slim and standard disk models. 
The Thomson optical depth can be estimated by the X-ray spectrum. We use these spectral templates as a function of the disk luminosity $L_{\rm disk}$ and $M_{\rm BH}$. 
Note that the relationship between the observed X-ray and optical/UV fluxes is generally non-trivial in the TDE case~\citep[e.g.,][]{Auchettl:2016qfa}. The disk state would change as time, and early-time emission may originate from the super-Eddington accretion. Also, the X-ray and UV emission can be obscured and reprocessed by the TDE debris. 

Protons may be accelerated to relativistic energies by plasma turbulence ~\citep[e.g.,][]{Lynn:2014dya,Comisso:2018kuh,Kimura:2018clk,Wong:2019dog} and/or magnetic reconnections~\citep[e.g.,][]{Zenitani:2014hea, 2015MNRAS.450..183S,2017ApJ...850...29R,2018MNRAS.473.4840W,2019ApJ...880...37P}.
For example, the stochastic acceleration time scale is $t_{\rm acc}\approx \eta_B{(c/V_A)}^2(H/c){(\varepsilon_p/eBH)}^{2-q}$, where $\varepsilon_p$ is the proton energy, $H$ is the coronal scale height, $V_A$ is the Alfv\'en velocity, $q\sim1.5-2$ is the spectral index of turbulent power spectrum, and $\eta_{B}$ is the inverse of the turbulence strength~\citep[e.g.,][]{Dermer:1995ju,Dermer:2014vaa}. 
The stochastic acceleration process is known to be slower than the diffusive shock acceleration, which can compete with various cooling and escape processes. 
For high Eddington-ratio objects (e.g., smaller SMBHs for a given $L_{\rm disk}$), the Bethe-Heitler pair production ($p\gamma\rightarrow pe^+e^-$) becomes the most important proton cooling process because of copious disk photons, and often determines the proton maximum energy~\citep{Murase:2019vdl}. 
CRs that are subject to efficient Bethe-Heitler cooling can still produce neutrinos via photomeson production, but the neutrino flux is significantly suppressed. 
For low Eddington-ratio objects (e.g., larger SMBHs for a given $L_{\rm disk}$), while the maximum energy is often limited by particle escape (either diffusion or infall), $pp$ inelastic collisions are more likely to be responsible for high-energy neutrino production. 
However, we stress that both $p\gamma$ and $pp$ contributions are important in the corona model. The $pp$ effective optical depth is given by~\citep{Murase:2019vdl}
\begin{eqnarray}
f_{pp}\approx n_p \kappa_{pp}\sigma_{pp}R\left(\frac{c}{V_{\rm fall}}\right)\sim3\left(\frac{\tau_T}{0.5}\right)\alpha_{-1}^{-1}\left(\frac{R}{30 R_{\rm S}}\right)^{1/2}
\end{eqnarray}
where $\sigma_{pp}\sim4.5\times{10}^{-26}~{\rm cm}^2$ is the $pp$ cross section, $\kappa_{pp}\sim0.5$ is the proton inelasticity, $V_{\rm fall}=\alpha V_K$ is the infall velocity, and $\tau_T=\sigma_T n_p H$ is the Thomson optical depth. The system is typically calorimetric in the sense that almost all CRs are depleted. 

To obtain CR spectra, with the code used in \cite{Kimura:2014jba,Murase:2019vdl,Kimura:2019yjo}, we solve the Fokker-Planck equation with terms for acceleration (momentum diffusion), cooling, escape, and injection, until a steady state is realized. In our model, since the outer disk radius is assumed to be larger than the emission region, the mass accretion rate is constant within its dynamical timescale, justifying the steady-state treatment.

As an illustrative example, we adopt parameters motivated by late-time observations of AT2019dsg.
\icnu was observed at $\sim10^7$~s post-discovery of \tde, at which the bolometric optical and UV luminosity was $L_{\rm OUV}\sim3\times10^{43}{\rm~erg~s^{-1}}$~\citep{Stein:2020xhk}. For $L_{\rm disk}=L_{\rm OUV}=10^{43.5}~{\rm egr}~{\rm s}^{-1}$, the effective temperature, $\sim3$~eV, is consistent with the observed temperature, $T_{\rm OUV}=10^{4.6}$~K. Correspondingly, we have $\lambda_{\rm Edd}\sim0.03M_{\rm BH,7}^{-1}$ and $L_X\sim3\times10^{42}\rm~erg~s^{-1}$. (Note that our results on the neutrino flux are unaffected even if lower X-ray luminosities are used.)
We consider two indicative values of the SMBH mass: $M_{\rm BH,7}=1$ and 3, which are compatible with $M_{\rm BH,7}\sim3$ implied from the buldge mass estimate~\citep{Stein:2020xhk}.
We adopt $R=30R_S$, $\alpha=0.1$, $\beta=1$, $q=5/3$, and $\eta_B=10$. Given these parameters, we can estimate the target photon field and hydrodynamical quantities in the coronae (see \citealt{Murase:2019vdl} for details). 

\begin{figure}[t]
\includegraphics[width=\linewidth]{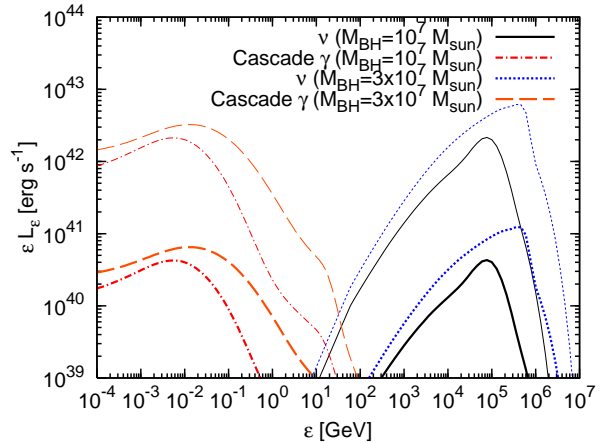}
\caption{Differential neutrino and cascade gamma-ray luminosities for the TDE corona model presented in Section~\ref{sec:corona}. We show results for $L_{\rm disk}=L_{\rm OUV}=10^{43.5}~{\rm erg}~{\rm s}^{-1}$ with $M_{\rm BH,7}=1$ and $M_{\rm BH,7}=3$ (see inset legend). 
The ratio of the CR pressure to the thermal pressure is set to 50\% for the most optimistic case (thin curves) and $1$\% for the modest case (thick curves).
\label{fig:corona}
}
\vspace{-1.\baselineskip}
\end{figure}

The results for our corona model are shown in Figure~\ref{fig:corona}. 
The neutrino spectrum shows a cutoff at $\varepsilon_\nu\sim100$~TeV ($\sim500$ TeV) for $M_{\rm BH,7}=1$ ($M_{\rm BH,7}=3$), respectively. This is because the CR spectrum is strongly suppressed at $\varepsilon_p\sim1$~PeV ($\sim5$ PeV) for $M_{\rm BH,7}=1$ ($M_{\rm BH,7}=3$), due to efficient photohadronic interactions with UV photons. 
In our cases, $\ll100$~TeV neutrinos mainly originate from $pp$ interactions but the photomeson production is also important for $\gtrsim100$~TeV neutrinos. 
We show the results for two different values of the ratio of the CR pressure to thermal pressure, namely 1\% and 50\%. The former is consistent with the corona model that explains the diffuse neutrino flux of in the 10-100~TeV energy range~\citep{Murase:2019vdl}. The latter can be regarded as an upper limit placed by the dominance of CR-induced radiation.

High-energy gamma-rays accompanied by the high-energy neutrino signal are absorbed by disk and coronal photons through the $\gamma \gamma\rightarrow e^+e^-$ pair production process. The pairs are eventually reprocessed to lower energies via either inverse-Compton or synchrotron emission, and escape from the source mostly as MeV photons. In Figure~\ref{fig:corona}, gamma-ray spectra up to 100~GeV energies are shown.
Note that we do not consider possible further reprocessing outside the corona due to Compton downscattering in the TDE debris. Although it depends on the details of the fate and geometry of the TDE debris and disk wind, the outer optical depth should decrease with time, so the gamma-ray signal can be a promising target for MeV gamma-ray telescopes. 

\subsection{RIAF model}\label{sec:riaf}
At early times, accretion is expected to take place through a slim or geometrically-thick disk and later a standard disk. The disk state will eventually change to a RIAF~\citep{1994ApJ...428L..13N,Yuan:2014gma} when the accretion rate decreases sufficiently. 
If the disk evolution\footnote{Here, $R_d$ is the initial disk radius of the standard disk phase. If the slim disk produces strong outflows, $R_d$ may be close to the circularization radius.} follows the solution of \cite{1990ApJ...351...38C}, the transition time is estimated to be $t_{\rm RIAF}\approx(\eta_{\rm rad}\dot{m}_{\rm crit})^{-16/19}t_{\rm vis}\simeq5.9\times10^8{\rm~s}\,\alpha_{-1}^{-51/19}M_{\rm BH,7}^{35/19}\mathcal{H}_{-1}^{-2}(R_d/10 R_S)^{3/2}$. 
In this case, the RIAF phase appears almost 20 yrs after the time of peak luminosity. However, the transition time may be shorter if the SMBH mass and the disk viscosity are higher than assumed here. Alternatively, the initial fallback rate can be suppressed by partial disruption or marginally hyperbolic orbits~\citep{Hayasaki:2018hxv}, or perhaps inefficient circularization. 
Outflows during the viscous evolution phase \citep[e.g.,][]{Nomura:2018vez} also help reduce the accretion rate.  
This idea is supported by observations of some TDEs that showed a plateau in their light curves~ \citep[e.g.,][]{Leloudas:2016rmh,vanVelzen:2018dlv,Wevers:2019dxz,Hung:2020jqz}. 

In RIAFs, the bulk of the accretion flow consists of collisionless plasma, in which non-thermal proton acceleration may operate. Here, we follow the formalism in \citet{Kimura:2019yjo,Kimura:2020thg} to calculate neutrino and gamma-ray emission. We estimate the neutrino luminosity at the time of the state transition. The mass accretion rate in the RIAF changes with the viscous timescale of the outer accretion disk, which can be as long as $t_{\rm vis}\simeq1.4\times10^8{\rm~s}~\alpha_{-1}^{-1}M_{\rm BH,7}\mathcal{H}_{-1}^{-2}(R_d/100 R_S)^{3/2}$. Since this is longer than the typical observed timescale of the TDEs, we will estimate the neutrino number assuming a constant neutrino flux for 1~yr in Section~\ref{sec:AT2019dsg}. We use the critical accretion rate of the state transition of $\dot{m}=\dot{m}_{\rm crit}\approx3\alpha^2\simeq0.03\alpha^2_{-1}$~ \citep{Mahadevan:1997qz,Xie:2012rs} .

To estimate the physical quantities, we use the analytic expressions from \citet{Kimura:2019yjo}, which are in rough agreement with global MHD simulations \citep[e.g.,][]{McKinney:2006tf,Ohsuga:2011jk,Narayan:2012yp,Suzuki:2013rka}.
As in the corona model (although the plasma beta is expected to be much larger), we consider particle acceleration by plasma turbulence and/or magnetic reconnection, and solve the Fokker-Planck equation where the acceleration is determined by two parameters, $\eta_B$ and $q$. As the escape process, we only consider infall escape and ignore diffusive escape, because the diffusive motion in vertical and radial directions are inefficient in RIAFs~\citep{Kimura:2016fjx,Kimura:2018clk}. If target photons are provided by thermal electrons heated by Coulomb collisions, we have $L_{\rm bol}\approx\eta_{\rm rad}\dot{m}_{\rm crit}L_{\rm Edd}(\dot{m}/\dot{m}_{\rm crit})^2$~\citep{Mahadevan:1997qz}. The electrons emit soft photons through synchrotron and Comptonization processes, which are calculated by the method in \citet{Kimura:2014jba}. The electron temperature is determined such that the electron cooling rate balances the heating rate. For a mass accretion rate close to $\dot{m}_{\rm crit}$, the photon spectrum is so hard that Bethe-Heitler pair production is sub-dominant unless we consider other sources of the target photon field. 

\begin{figure}[t]
\includegraphics[width=\linewidth]{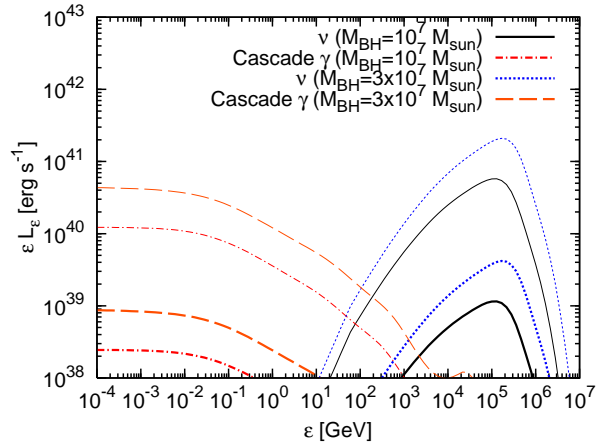}
\caption{Differential neutrino and gamma-ray luminosities expected in the TDE RIAF model. 
We use $M_{\rm BH,7}=1$ and 3 for $\dot{m}=\dot{m}_{\rm crit}\approx0.03\alpha_{-1}^2$, which correspond to $L_{X}\sim7\times10^{41}~{\rm erg}~{\rm s}^{-1}$ and $L_{X}\sim3\times10^{41}~{\rm erg}~{\rm s}^{-1}$, respectively. The ratio of the CR pressure to the thermal pressure is set to 50\% for the most optimistic case (thin curves) and 1\% for the modest case (thick curves).
\label{fig:RIAF}
}
\vspace{-1.\baselineskip}
\end{figure}

For the RIAF model, we adopt $\alpha=0.1$, $\beta=10(>1)$, $R=10R_S$, $\eta_B=10$, and $q=5/3$~\citep[see model A of][]{Kimura:2020thg}. 
We use $M_{\rm BH,7}=1$ and 3, and $\dot{m}=\dot{m}_{\rm crit}\approx0.03\alpha^2_{-1}$.
The resulting neutrino and gamma-ray spectra are shown in Figure~\ref{fig:RIAF}. 
The neutrino emission mainly comes from inelastic $pp$ interactions. In general, the neutrino spectrum in the corona model is more modulated because the Bethe-Heitler process and photomeson production are not negligible in the corona model and become dominant for luminous objects~\citep{Murase:2019vdl}. 
The GeV-TeV spectrum of gamma-rays accompanied by the neutrinos is suppressed by the two-photon pair annihilation, so RIAFs serve as gamma-ray--hidden neutrino sources.

In the RIAF case, CR acceleration is limited by escape and inelastic $pp$ interactions, resulting in the spectral softening around $\varepsilon_\nu\sim10^4$ GeV. 
The spectral softening is slow due to the weak energy dependence of $pp$ and infall losses. 
The photomeson production can be effective only at energies higher than the maximum energy~\citep{Kimura:2019yjo}, making a sharp cutoff in the neutrino spectrum.  
Neutrinos are still produced predominantly through $pp$ interactions. The effective $pp$ optical depth for $\dot{m}=\dot{m}_{\rm crit}\approx3\alpha^2$ is~\citep{Kimura:2019yjo}
\begin{eqnarray}
f_{pp}\approx\frac{24\sigma_{pp}\kappa_{pp}}{\sigma_T}\sim0.8,
\end{eqnarray}
which is independent of parameters such as $\alpha$, $\beta$, $R$, and $M_{\rm BH}$, and the system is almost calorimetric. 

The neutrino luminosity is an order of magnitude lower than that in the corona model because of the lower accretion rate, which translates to a lower CR production rate. The total luminosity is limited by $\dot{m}_{\rm crit}L_{\rm Edd}\sim4\times10^{43}~M_{\rm BH,7}\alpha_{-1}^2\rm~erg~s^{-1}$. 
The all-flavor neutrino luminosity for a given $M_{\rm BH}$ can be written as
\begin{eqnarray}
&\varepsilon_\nu& L_{\varepsilon_\nu} \approx \frac12 f_{pp}\frac{\varepsilon_p L_{\varepsilon_p}}{\mathcal{R}_{\rm CR}}\nonumber\\
& & \lesssim10^{41}~{\rm~erg~s^{-1}}~\alpha_{-1}^2\left(\frac{\mathcal{R}_{\rm CR}}{3}\right)^{-1}\left(\frac{\eta_{\rm CR}}{0.025}\right)M_{\rm BH,7},\,\,\,\,\,\,\,\,
\label{eq:LnuRIAF}
\end{eqnarray}
where $\mathcal{R}_{\rm CR}\ge1$ is a bolometric correction factor and $\eta_{\rm CR}\equiv \int L_{\varepsilon_p}d\varepsilon_p/\dot{M}c^2$ is the energy conversion factor of accretion power to CR protons. 
The released gravitational energy is shared by the bulk motion, thermal protons, and CRs and other emission. 
In our model for a given $R$, in the limit that all CRs are depleted for radiation, the virial theorem implies that the CR luminosity is limited by $GM_{\rm BH}\dot{M}/(2R)=(\dot{M}c^2/40)\,(R/10R_S)^{-1}$, leading to $\eta_{\rm CR}<(1/40)\,(R/10R_S)^{-1}$.
The neutrino luminosity for the most optimistic case shown in Figure~\ref{fig:RIAF} is close to the upper limit by Equation~(\ref{eq:LnuRIAF}).

\section{Hidden wind model}
\label{sec:shock_model}
TDEs originate from the disruption of a star by a SMBH. While the bound material has elliptical trajectories with large apocenter distances, the unbound material has hyperbolic orbits. The orbits of tidal streams are highly eccentric, and the most bound stellar debris has an orbit with $a_{\rm min}$ (see Equation~\ref{eq:amin}). 
It is natural that the circularization involves shock dissipation, and the returning flow may collide with the streaming inflows~\citep[e.g.,][]{Kochanek:1993cm,Shiokawa:2015iia,Guillochon:2015qfa,Jiang:2016igx,Hayasaki:2018hxv,Lu:2019hwv,Andalman:2020tjr}. 
It has been suggested that the consequent shock heating powers the observed optical/UV emission~\citep{Piran:2015gha,Svirski:2015fma}. 
The available energy for such stream crossing shocks and subsequent secondary shocks is estimated to be
\begin{eqnarray}
{\mathcal E}_{\rm scs}&\approx&\frac{GM_{\rm BH}\eta_{\rm fb}(M_*/2)}{2a_{\rm min}}\nonumber\\
&\simeq&9.4\times{10}^{50}~{\rm erg}\, f_{T-1.1}^{-1/3}\eta_{\rm fb}M_{*}^{2/3+\xi}M_{\rm BH,7}^{1/3}.
\label{gravapo}
\end{eqnarray}
The relative velocity between shocking streams at the apocenter region is the Keplerian velocity,
\begin{eqnarray}
V_{\rm scs}&\approx&\sqrt{\frac{GM_{\rm BH}}{a_{\rm min}}}\nonumber\\
&\simeq&1.4\times{10}^{9}~{\rm cm}~{\rm s}^{-1} f_{T,-1.1}^{-1/6}M_{*}^{-1/6+\xi/2}M_{\rm BH,7}^{1/6}. \,\,\,
\label{velapo}
\end{eqnarray}
CRs could be accelerated by these shocks given that the shock is ``unmediated'' by radiation (see below). 

A significant fraction of the shocked debris can be unbound as an outflow for massive SMBHs~\citep{Lu:2019hwv}, which may be responsible for soft X-ray attenuation, reprocessed optical/UV emission that is observed, and radio emission by sub-relativistic flows with $\sim(0.01-0.1)c$. 
The remaining fraction may form a geometrically-thick disk, whose radius is much larger than $R_T$, and a quasi-spherical weakly-bound debris~\citep{Loeb:1997dv,Coughlin:2013lva,Sadowski:2015jor,Bonnerot:2019yjb,Andalman:2020tjr}. 
Such ``TDE debris'' is schematically depicted in Figure~\ref{fig:TDE}.   

Another possibility is the dissipation caused by sub-relativistic disk-driven winds. Details will depend on the fate of the disk especially in the inner region. In addition to disk accretion, a fraction of TDE debris would accrete onto a SMBH via the funnel~\citep{Sadowski:2015jor,Bonnerot:2019yjb}. The wind is expected to be launched from the vicinity of the SMBH by radiation from a slim or geometrically-thick disk~\citep[e.g.,][]{Strubbe:2009qs,Metzger:2015sea}, line emission~\citep{Miller:2015jda}, or magnetic dissipation, which may further interact with the TDE debris and streams mentioned above. Additional dissipation might occur via internal shocks because the wind base may be variable on $t_{\rm diss}\approx\alpha^{-1}\Omega_K^{-1}\simeq2.2\times10^5~{\rm s}~{(R/30R_S)}^{3/2}M_{\rm BH,7}$. 
Particle acceleration associated with magnetic dissipation in the magnetized wind has also been considered~\citep{Xiao:2016man}. 
The wind velocity around the classical circularization radius at $\sim2R_T$ is estimated to be~\citep[e.g.,][]{Metzger:2015sea}
\begin{eqnarray}
V_w\approx\sqrt{\frac{\eta_{\rm in}GM_{\rm BH}}{2R_T}}&\simeq&2.6\times{10}^{9}~{\rm cm}~{\rm s}^{-1}~\eta_{\rm in,-1}^{1/2}\nonumber\\
&\times& f_{T,-1.1}^{-1/12}M_{\rm BH,7}^{1/3}M_*^{\xi/2-1/3}, \,\,\,\, 
\label{velw}
\end{eqnarray}
which can be larger than Equation~(\ref{velapo}). 
Assuming that most of the fallback material blown out by the wind and the wind is so optically thick that radiation losses are negligible, the kinetic energy of the wind-driven TDE debris is estimated to be
\begin{eqnarray}
{\mathcal E}_w\approx\frac{1}{2}\frac{\eta_{\rm fb}M_*}{2}V_w^2&\simeq&3.4\times 10^{51}~{\rm erg}~\eta_{\rm in,-1}\eta_{\rm fb}M_*^{1/3+\xi}\nonumber\\
&\times& f_{T,-1.1}^{-1/6}M_{\rm BH,7}^{2/3},
\label{kinw}
\end{eqnarray}
and can be somewhat larger than the energy given by Equation~(\ref{gravapo}) due to the higher escape velocity in the inner disk. 
Note that the kinetic energy is comparable to that of powerful Type IIn SNe such as SN 2010jl although the velocities are different~\citep[see][and references therein]{Murase:2018okz}.

Hereafter we assume that CRs are accelerated by high-velocity winds embedded in the TDE debris or possibly shocks induced by stream-stream collisions, and consider hadronic interactions by escaping CRs in the wind bubble and the debris material. 
Note that the debris near the apocenter would be optically thick especially at early times. The Thomson optical depth at $R$ is estimated to be
\begin{equation}
\tau_T^{\rm deb}\approx\frac{3\sigma_T M_{\rm deb}}{4\pi R^2m_p}\simeq94~\left(\frac{M_{\rm deb}}{0.5M_{*}}\right)R_{15}^{-2},
\end{equation}
where $M_{\rm deb}\lesssim\eta_{\rm fb}M_*/2$ is the debris mass.  
If the TDE debris is bound (for $R\lesssim a_{\rm min}$), one may use $f_{pp}\approx \kappa_{pp}\sigma_{pp}n_p ct_{\rm diff}$, where $t_{\rm diff}$ is the CR diffusion time. On the other hand, the shock-driven and/or wind-driven unbound debris may homologously expand with $V_{\rm deb}\sim V_{\rm scs}$ or $V_w$.
As long as the CR diffusion time is longer than the expansion time, the effective $pp$ optical depth for CRs interacting with the unbound TDE debris is given by 
\begin{equation}
f_{pp}\approx \frac{3 \kappa_{pp}\sigma_{pp}M_{\rm deb}}{4\pi R^2m_p}\left(\frac{c}{V_{\rm deb}}\right)\sim1~\left(\frac{M_{\rm deb}}{0.5M_{*}}\right)R_{16}^{-2} V_{\rm deb,9}^{-1}.
\end{equation}
Thus, the CRs can be depleted once they leave the wind and interact with the debris. 
Electrons may also be accelerated around the wind termination radius~\citep[][for discussion in the neutron star merger case]{Murase:2017snw}, but CR-induced hadronic emission can be dominant due to high efficiencies of photomeson production and inelastic $pp$ collisions. 

The differential neutrino luminosity is estimated by
\begin{equation}
\varepsilon_\nu L_{\varepsilon_\nu}\approx\frac{3K}{4(1+K)}f_{\rm mes}f_{\rm geo}\frac{\epsilon_{\rm CR}L_{w}}{\mathcal{R}_{\rm CR}},
\label{eq:Lnuwind}
\end{equation}
where $f_{\rm mes}\approx{\rm min}[1,f_{pp}+f_{p\gamma}]$ is the meson production efficiency, which can be either by hadronuclear ($pp$) or photomeson production ($p\gamma$) process, and $K=1$ and $K=2$ for $p\gamma$ and $pp$ interactions, respectively. Also, $\epsilon_{\rm CR}$ is the energy fraction carried by CRs and $L_{w}$ is the wind luminosity. A few remarks about Equation~(\ref{eq:Lnuwind}) follow. 
First, the bolometric correction in the CR spectrum $\mathcal{R}_{\rm CR}$ should not be ignored. Assuming an $\varepsilon_p^{-2}$ spectrum and for $\varepsilon_p^{\rm max}\sim10^{7}$~GeV, we have $\mathcal {R}_{\rm CR}=\ln(\varepsilon_p^{\rm max}/\varepsilon_p^{\rm min})\sim16$. Steeper CR spectra lead to larger values.  
Second, it is natural to expect that the debris is not spherical and a fast wind or jet would be launched preferentially toward the polar region~\citep[e.g.,][]{Sadowski:2015jor,Dai:2018jbr}. 
In the case of \tde, the radio emission could originate from a mildly-relativistic outflow with a large opening angle powered by the central engine \citep{Stein:2020xhk}.
In this case, depending on the solid angle of the surrounding debris, only a fraction of the CRs may experience $pp$ interactions (described by the factor $f_{\rm geo}\equiv\Delta\Omega/(4\pi)$; see also \citet{Murase:2018okz} for discussion in the case of non-spherical target material), while the remaining CRs will escape from the wind. 
Third, $f_{pp}\gtrsim1$ decreases with time and defines the critical radius $R_{pp}$ at which $f_{pp}=1$.   
Even if the wind luminosity is constant (different from the standard value of $5/3$), the debris swept by the wind may eventually accelerate it to $V_w$. Higher values of $V_{\rm deb}$ reduce $f_{pp}$ for a given time. 
The above considerations imply that our calculations should be regarded as optimistic.  

\begin{figure}[t]
\begin{center}
\includegraphics[width=0.9\linewidth]{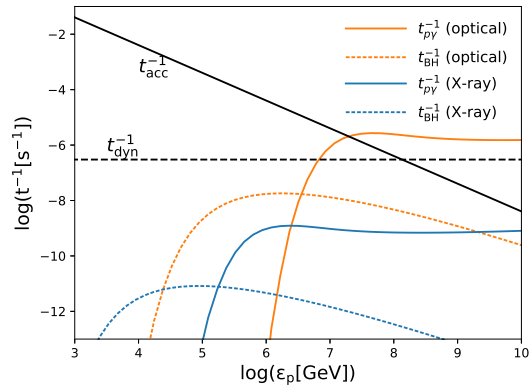}
\caption{Cooling (colored lines), acceleration (solid black line), and dynamical (dashed black line) rates of protons in the wind region. The dissipation radius is set to $R_{\rm diss}={10}^{16}$~cm, where the shock velocity and magnetic field strength are set to $V_w=0.1c$ and $B_w=3$~G. We use gray body spectra with $L_{\rm OUV}=10^{43.5}~{\rm erg}~{\rm s}^{-1}$ and $T_{\rm OUV}=10^{4.6}$~K, and $L_{X}=10^{42.5}~{\rm erg}~{\rm s}^{-1}$ and $T_{X}=10^{5.9}$~K. 
\label{fig:time}
}
\vspace{-1.\baselineskip}
\end{center}
\end{figure}

One of the necessary conditions for conventional shock acceleration to be efficient is the radiation constraint, $\tau_T\lesssim c/V$~\citep[see][for details in the case of non-relativistic shocks]{Murase:2010cu,Murase:2018okz}. 
Efficient particle acceleration does not occur when the shock is radiation mediated~\citep{Murase:2010cu}. As shown above, the Thomson optical depth of the TDE debris around the apocenter is expected to be large, so CR acceleration at the forward shock can occur only at late times; shock acceleration near the SMBH is also difficult~\citep{Hayasaki:2019kjy}. 
On the other hand, CR acceleration in the wind zone far from the launching region is easier, because the Thomson optical depth at $2R_T\ll R_{\rm diss}<R$, 
\begin{equation}
\tau_T^w\approx\frac{\sigma_T\dot{M}_w}{4\pi RV_wm_p}\simeq1.1\dot{M}_{w,26}R_{\rm diss,15}^{-1}\left(\frac{V_w}{0.1c}\right)^{-1},
\end{equation}
which is less than $c/V_w$ so the efficient CR acceleration may operate. Here, a spherically symmetric wind is assumed for simplicity. In this case, the magnetic field strength is estimated to be $B_w=\sqrt{2\epsilon_BL_w/(R_{\rm diss}^2 V_w)}\simeq2.5~{\rm G}~{(\epsilon_B/0.03)}^{1/2}L_{w,43.5}^{1/2}R_{\rm diss,16}^{-1}{(V_w/0.1c)}^{-1/2}$, implying that protons can be accelerated to $\sim10-100$~PeV energies (assuming the Bohm limit for the turbulence). In reality, CRs in the wind bubble are subject to various energy losses, and the photomeson production and Bethe-Heitler energy losses can be important. 
The energy loss rates which correspond to the conditions stated above are illustrated in Figure~\ref{fig:time}.  
Note that although the intrinsic X-ray luminosity is included assuming $L_X=0.1~L_{\rm OUV}$, X-rays do not affect CR energy losses for our parameters. 

Following the model described above, we numerically calculate the neutrino and gamma-ray emission. The method is similar to one used in the previous work~\citep{Murase:2018okz,Murase:2019vdl}. 
We assume that protons are accelerated at $R_{\rm diss}=R=10^{16}$~cm and the shock velocity is $V_w=0.1c$. 
Photohadronic interactions of escaping CRs within the wind region are simulated following~\cite{Zhang:2019htg}, in which the publicly available code {\sc CRPropa-3.0} is utilized~\citep{Batista:2016yrx}.
Target photon spectra are assumed to be two-component gray-body spectra, analogous to previous calculations for Type-IIn supernovae~\citep{Murase:2018okz}. 
The optical-UV and X-ray energy densities are implemented as $U_{\rm OUV}\approx [3(1+\tau_T)L_{\rm OUV}]/(4\pi R^2c)$ and $U_{X}\approx 3L_X/(4\pi R^2c)$, respectively. 
CRs leaving the wind diffuse in the TDE debris. The radiation luminosity is expected to be a fraction of the dissipation luminosity $\epsilon_{\rm rad}\sim0.2-0.5$. Following \cite{Murase:2018okz}, we normalize the CR luminosity by using the CR loading parameter $\xi_{\rm CR}=L_{\rm CR}/L_{\rm OUV}\sim0.1-1$~\footnote{\citet{Murase:2018okz} presented a phenomenological model to describe neutrino and gamma-ray emission taking into account the non-spherical geometry (i.e., $f_{\rm geo}<1$).
CR acceleration may operate after optical/UV photons break out~\citep{Murase:2010cu}. 
The radiation and CR luminosities are $L_{\rm OUV}=\epsilon_{\rm rad}f_{\rm geo}L_w$ and $L_{\rm CR}=\epsilon_{\rm CR}f_{\rm geo}L_w$, respectively.  Using the {\it Fermi}-LAT data, \citet{Murase:2018okz} obtained $\epsilon_{\rm CR}\lesssim 0.05 (\epsilon_{\rm rad}/0.25)$.}. 

\begin{figure}[t]
\includegraphics[width=\linewidth]{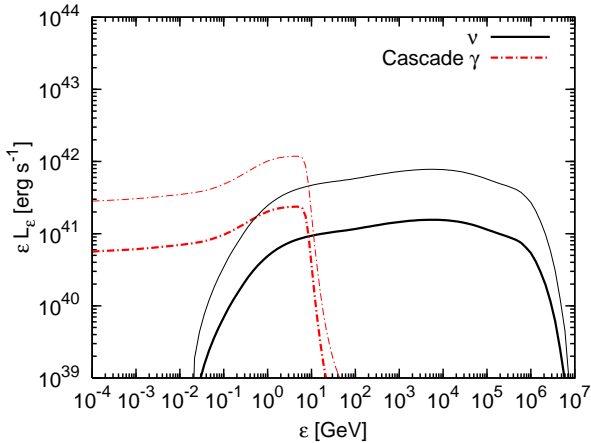}
\caption{Differential neutrino and cascade gamma-ray luminosities expected in the hidden wind model. 
Accelerated CRs interact with optical/UV (and X-ray) photons in the outflow and gas in the TDE debris (with $M_{\rm deb}=0.5~M_\odot$ and $V_{\rm deb}=10^9~{\rm cm}~{\rm s}^{-1}$ at $R=10^{16}$~cm). 
The CR loading parameter $\xi_{\rm CR}=L_{\rm CR}/L_{\rm OUV}$ is set to 1 for the optimistic case (thin curves) and $0.2$ for the modest case (thick curves).
\label{fig:shock}
}
\vspace{-1.\baselineskip}
\end{figure}

The resulting spectra of high-energy neutrinos and cascade gamma-rays (up to 100~GeV) in the hidden wind model are shown in Figure~\ref{fig:shock}. 
The system is calorimetric in the sense that CRs are mostly depleted via the photomeson production and inelastic $pp$ collisions, so that the neutrino energy spectrum is almost flat as in the injected CR spectrum (with modulations by the interplay of different cooling processes). The photomeson production is important in the PeV range, whereas the contribution of $pp$ interactions is dominant at lower energies. Although we add the intrinsic X-ray luminosity assuming $L_X=0.1L_{\rm TDE}$, our results on neutrino spectra are unaffected by the X-rays. In this sense the results are conservative.  
Electromagnetic cascades are developed mainly via two-photon pair annihilation and inverse-Compton emission, and subsequent regeneration processes lead to the prediction of gamma-rays below the GeV range. 
Interestingly, the spectral features of both neutrino and gamma-rays is similar to what was predicted for Type IIn SNe~\citep{Murase:2018okz,Petropoulou:2017ymv}. For gamma-rays, this is because the cutoff is caused by optical and UV photons. Further gamma-ray attenuation due to the Bethe-Heitler process in the debris is negligible for $V_{\rm deb}\lesssim 0.3 c~f_{pp}^{-1}$.

\section{Implications for AT2019dsg}
\label{sec:AT2019dsg} 
\subsection{Summary of observations}
The discovery of \tde triggered a follow-up campaign, much before the detection of \icnu. The details of the observations are presented in \citet{Stein:2020xhk}. 
Below we give a short summary. On April 9th 2019 ZTF reported the discovery of \tde as an optical transient of likely extragalactic origin~\citep{2019TNSTR.615....1N}. Spectroscopic observations of \tde with the extended-Public ESO Spectroscopic Survey for Transient Objects (ePESSTO+)~\citep{2019TNSAN..13....1N} classified it as a TDE. Radio follow-up observations first with the Arcminute Microkelvin Imager (AMI-LA)~\citep{2019ATel12798....1S} and later with the Enhanced Multi Element Remotely Linked Interferometer Network (e-MERLIN)~\citep{2019ATel12960....1P} revealed radio emission. \tde belongs to a rare type of TDEs which exhibit radio (non-thermal) emission, suggestive of particle acceleration to relativistic energies. 

UV emission from \tde was first detected by the \emph{Swift}-UltraViolet and Optical Telescope (UVOT) on 2019 May 17. By this time the UV emission was steadily dimming. The combined optical and UV emission of \tde was found to be well described by a blackbody spectrum of temperature $10^{4.59\pm0.02}$~K. The peak luminosity of \tde was estimated to be $10^{44.54\pm0.08}\rm erg~s^{-1}$, placing it in the top $10\%$ of known TDEs. Around the time of neutrino detection, the UV luminosity (a good proxy of the bolometric luminosity) was found to be $\sim 3\times10^{43}$~erg s$^{-1}$. 
Late time light curves are consistent with a plateau, which can be interpreted as the emission from an accretion disk~ \citep{Leloudas:2016rmh,vanVelzen:2018dlv,Wevers:2019dxz}.

\tde was detected in X-rays starting 37 days after its discovery first with the  \emph{Swift}-X-Ray Telescope (XRT)~\citep{2019ATel12777....1P} and later with the The Neutron star Interior Composition Explorer, (\emph{NICER}), and the X-ray Multi-Mirror Mission,  (\emph{XMM-Newton})~\citep{2019ATel12825....1P}. The detected X-ray emission was soft, as found in other X-ray candidate TDEs~\citep{2017ApJ...838..149A}. The X-ray spectrum of the \emph{XMM-Newton} observation was well described by an absorbed blackbody with a temperature of $10^{5.9}$~K and hydrogen column density (Galactic and intrinsic) of $N_{\rm H}\sim 4\times10^{20}$~cm$^{-2}$. The X-ray flux declined rapidly, falling below the detection threshold of \emph{Swift}-XRT within 60 days post-discovery, and therefore much before the detection of \icnu. A second \emph{XMM-Newton} observation performed on 2019 October 23 (i.e., after the detection of \icnu) yielded a deep upper limit of $9\times10^{-14}$~erg cm$^{-2}$~s$^{-1}$.

An analysis of data obtained with the \emph{Fermi} Large Area Telescope (LAT) in the direction of \tde revealed no significant ($<5\sigma$) gamma-ray emission from this source. The analyses performed spanned the period from 2019 April 4 to 2020 January 31, and several sub-periods~\citep{gcn25932,Stein:2019ivm}. Similarly, follow-up searches for TeV emission in response to the detection of \icnu with the High-Altitude Water Cherenkov Observatory (HAWC) and the First G-APD Cherenkov Telescope (FACT) resulted only in upper limits~\citep{gcn25936,gcn25946}. 

\subsection{Summary of model predictions}
In the previous sections, we provided several models for neutrino and gamma-ray emission from TDEs. We consider model-dependent implications, including the hidden jet model suggested in \cite{Senno:2016bso}, for \icnu below. 

\begin{figure}[t]
\includegraphics[width=\linewidth]{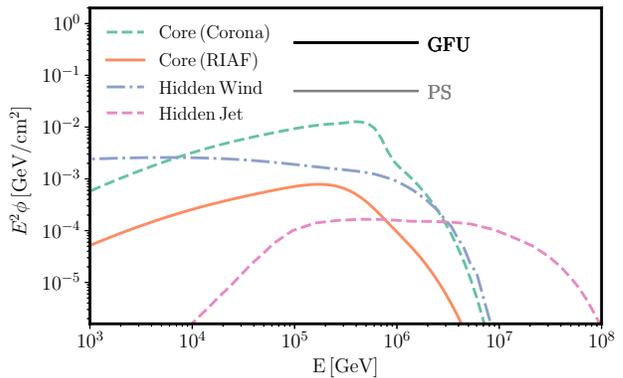}
\caption{Most optimistic all-flavor neutrino fluences expected for \tde in the corona (for $M_{\rm BH,7}=3$), RIAF (for $M_{\rm BH,7}=3$), hidden wind, and hidden jet scenarios. See also Figure~\ref{fig:TDE}. 
The horizontal lines show the fluence level needed to produce one neutrino in the GFU and PS channels respectively for an $E_\nu^{-2}$ neutrino spectrum. 
Note that the fluences are lower for the modest cases.
}
\label{fig:nu}
\end{figure}

Figure~\ref{fig:nu} summarises the most optimistic all-flavor neutrino fluences from the models considered for \tde in Sections~\ref{sec:core_models} and \ref{sec:shock_model}, for an assumed duration of one year after the discovery of \tde. We additionally show the prediction of a hidden jet model, previously studied by~\cite{Senno:2016bso}. This case is optimistic because $\epsilon_{\rm CR}=1$ (i.e., almost all the jet energy goes to CRs in the on-axis TDE) is used and we further push the neutrino fluence by considering $t_{\rm dur}=3\times 10^{6}$~s (for details, see section~\ref{sec:hidden-jet}).
The horizontal lines show the all-flavor neutrino flux that \tde must produce in order to produce one muon neutrino in IceCube. It is evident that all models fall short of producing the required flux to expect one event, but the most promising model is the Core (Corona) model.

We additionally estimate the number of muon and anti-muon neutrinos expected to be observed with IceCube as
\begin{equation}
{\mathcal N}_{\nu_{\mu}}~=~\int_{E_{\nu_{\mu},{\rm min}}}^{E_{\nu_{\mu},{\rm max}}} {\rm d} E_{\nu_{\mu}} A_{\rm eff}(E_{\nu_{\mu}},\delta) \phi_{\nu_{\mu}},
\end{equation}
where $E_{\nu,{\rm min}}$=100~TeV and $E_{\nu,{\rm max}}=$2~PeV, given the energy range where one expects $90\%$ of neutrinos in the GFU channel at the declination $\delta$ of \tde, $\phi$ is the muon neutrino fluence, and $A_{\rm eff}$ is the effective area. 
We also consider the two effective areas representing the real-time alert event selection and point-source event selection at the declination of \tde. The effective area of the IceCube Point Source (PS) analysis is taken from \citet{Aartsen:2018ywr}, whereas we use the area of the Gamma-ray Follow-Up (GFU) selection~\citep{Blaufuss2019ICRC} for the IceCube alert analysis. The latter is smaller than the PS effective area, so the neutrino fluence level inferred from the PS analysis allows of more conservative discussion given the population bias.
Table~\ref{tab:nu} gives the estimated number of expected neutrinos in each of the models we studied. 
We discuss the implications of these results for each model separately below. 

\begin{table}[]
\centering
\begin{tabular}{ccc}
\hline
Model & \multicolumn{2}{c}{${\mathcal N}_{\nu_{\mu}}(>100 \, \rm TeV)$} \\
 & Point Source & GFU \\
\hline
Core (Corona)  &  $9\times10^{-2}$  &  $1\times10^{-2}$  \\
Core (RIAF) &  $3\times10^{-3}$  &  $3\times 10^{-4}$  \\ 
Hidden Wind  &  $9\times10^{-3}$  &  $1\times10^{-3}$  \\ 
Hidden Jet &  $1\times10^{-3}$  &  $3\times10^{-4}$
 \\ 
 \hline
\end{tabular}
\caption{Maximum expected number of muon and antimuon neutrinos with energy exceeding 100 TeV in the PS and GFU channels for the models studied in Sections~\ref{sec:core_models}-\ref{sec:shock_model}. The hidden-jet model is discussed in Section~\ref{sec:hidden-jet}. 
Note that all the values are for the most optimistic cases, and we expect smaller values for the modest cases.  
}
\label{tab:nu}
\end{table}

\subsubsection{Core models: possible}
We calculate the expected number of muon neutrinos, by optimistically assuming an integration time of $\Delta T=1$~yr. 
For the conditions assumed in the corona model we obtain ${\mathcal N}_{\nu_{\mu}}=0.09\,(0.01)~\rm yr^{-1}$ with the Point Source (GFU) effective area\footnote{Henceforth, the number of neutrinos enclosed in the parenthesis refers to the GFU effective area.}. 
Even the most optimistic expected number of neutrinos is less than unity, but there is still a $\lesssim 10\%~(1\%)$ chance to detect one neutrino taking into account model uncertainties. This expectation value leads us to conclude that the neutrino could in principle have been produced by \tde, if the physical conditions of the core model were in place, and can be interpreted as an upward statistical fluctuation.

Plateaus in optical/UV light curves are often interpreted by the emission from accretion disks~ \citep{Leloudas:2016rmh,vanVelzen:2018dlv,Wevers:2019dxz}. Intriguingly, \tde showed the plateau around the detection time of \icnu~\citep{Stein:2020xhk}. 
Note that the X-ray luminosity used in the model is consistent with the upper limit around the time of the neutrino detection~\citep{Stein:2020xhk}. There could be a possible tension with a late-time limit, but the details depend on the time evolution of the disk emission and the level of obscuration by the TDE debris.
The neutrino may be associated with the formation of the disk-corona structure, which implies that high-energy neutrinos can be used as a probe of the TDE disk that is difficult to probe by electromagnetic observations. 
In the corona model, the production of $\gtrsim100$~TeV neutrinos is allowed for sufficiently high-mass SMBHs (with low Eddington ratios). TDEs with typical optical/UV luminosities or smaller SMBHs predict lower-energy neutrinos with lower neutrino fluxes. This could explain why the neutrino production is accompanied by \tde-like luminous TDEs that are accompanied by powerful radio-emitting outflows.   

However, the CR pressure with 50\% of $GM_{\rm BH}m_pn_p/(3R)$ (that is the original thermal pressure for the virial temperature) is rather extreme. (In this limit, the non-thermal pressure is given by $GM_{\rm BH}m_pn_p/(6R)$, which corresponds to non-thermal energy equal to half of the gravitational binding energy without leaving room for thermal particles, although in the corona model additional energy can in principle be supplied by the disk.)
Although it cannot be excluded by the observations, this is a very strong energetic requirement. However, lower neutrino fluences with more conservative CR normalization can also be consistent with the observation of one neutrino from the entire (known) TDE population under the assumption that the relevant conditions exist in all ZTF observed TDEs (see discussion on the population bias in \citealp{Stein:2020xhk,Strotjohann:2018ufz}).

The accompanying gamma-rays should be significantly attenuated in the GeV-TeV range. The cascade gamma-rays are well below the Fermi upper limit, which is $\sim{\rm a~few}\times10^{43}~\rm erg~s^{-1}$ in the $0.1-800$~GeV energy range assuming a photon index of $\Gamma_\gamma=2$. The corona model is consistent with the gamma-ray upper limits obtained with the \emph{Fermi}-LAT and HAWC. 

The RIAF model is less consistent with the observation of a neutrino from \tde, as the expectation is ${\mathcal N}_{\nu_{\mu}}=3\times10^{-3}(3\times 10^{-4})~\rm yr^{-1}$, with the Point source (GFU) effective area. 
With extremely optimistic parameters in Equation~(\ref{eq:LnuRIAF}), one could increase the neutrino luminosity by considering inner disk regions, but it is still challenging to account for the observation of one neutrino event.
On other other hand, from the observations of \tde, the disk luminosity is estimated to be $\sim10^{43}\rm~erg~s^{-1}$ 100 days after the peak. Interestingly, with $M_{\rm BH}\sim3\times10^7M_\odot$, this luminosity corresponds to the Eddington ratio of $\sim0.003$, which is compatible with the critical luminosity at which the state transition is expected for $\alpha=0.1$. Although the state transition time scale is typically expected to be much longer than 1 yr, the RIAF phase may start significantly earlier if the majority of the stellar debris is unbound or most of the fallback material is ballistically swallowed by a SMBH ~\citep[e.g.,][]{Svirski:2015fma,Hayasaki:2018hxv}. 
The fallback and circularization mechanisms of the disrupted star have been debated for a long time, and further studies are necessary. 
Also, RIAFs near the critical accretion rate emit most of the heating energy as MeV gamma-rays~\citep{Kimura:2020thg}, and hence, our RIAF model cannot explain the observed UV photons. They should be attributed to a different emission site, such as stream collisions or the outer accretion disk. The observed UV photons do not significantly affect the neutrino emission in our RIAF model, while they suppress the gamma-rays above $\sim10$ GeV. 


If one considers MADs~\citep{1974Ap&SS..28...45B,Narayan:2003by}, the reconnection layer at the disk-outflow boundary may have a plasma beta of $\beta\lesssim1$ and the magnetization parameter may be as low as $\sigma\gtrsim1$~\citep[e.g.,][]{Ball:2017bpa,Ripperda:2020bpz}, and CR acceleration through reconnections could be more efficient~\citep[e.g.,][]{2015MNRAS.450..183S, 2018MNRAS.473.4840W}.
However, the neutrino luminosity from RIAFs would still be limited by Equation~(\ref{eq:LnuRIAF}), which is also applicable to RIAF MADs. Hence, it would be challenging to explain the observed neutrino flux as long as we consider the RIAF regime. The flux can be enhanced for $M_{\rm BH}\gtrsim10^8\rm~M_\odot$, but TDEs are not expected to occur for such heavy SMBHs (due to $R_T\lesssim R_{\rm S}$). We do not discuss MADs in super-Eddington phase \citep{Hayasaki:2019kjy}, which is beyond the scope of this paper.

\subsubsection{Hidden wind model: challenging}
For the neutrino spectrum shown in Figure~\ref{fig:shock}, we calculate the number of muon neutrinos, by assuming an integration time of $\Delta T=1$~yr. Here we adopt a constant wind luminosity \citep[see Fig.~2 in][]{Stein:2020xhk}, although it would be optimistic to extrapolate it to one year because the accretion rate decreases as time. 
The impact of possible time dependence would also be small because the observational time is not far from $t_{\rm fb}$ for $M_{\rm BH}\gtrsim10^7~M_\odot$ and the corresponding CR energy input already reaches ${\mathcal E}_{\rm CR}=10^{51}$~erg).
Note that the differential neutrino energy per flavor is at most $\sim{10}^{49}$~erg because of the CR bolometric correction, ${\mathcal R}_{\rm CR}\sim20$. 

We obtain ${\mathcal N}_{\nu_{\mu}}=9\times10^{-3}\, (1\times10^{-3})~\rm yr^{-1}$, which implies that the probability to observe one muon neutrino would be at most $\sim0.1-1$\%. Such a small value could still be consistent with the observation of \icnu taking into account large population bias~\citep{Strotjohann:2018ufz}. 
But the consistency between the theoretical expectation and neutrino observation in this model is not compelling at this point, since there are several factors that can easily reduce the resulting neutrino fluence (e.g., $f_{\rm geo}<1$, $f_{pp}<1$ at late times due to the expansion), which have not been considered here.  

Nevertheless, the hidden wind model is attractive in several points. 
Radio observations of \tde are suggestive of the presence of a mildly relativistic outflow that could be either the disk-driven wind or the debris powered by the wind and/or stream crossing shocks, so CR acceleration looks promising.
The radius indicated by radio observations (a few~$\times{10}^{16}$~cm) around the neutrino detection time ($\sim150$~days after the discovery) is not far from the critical radius for escaping CRs to have efficient inelastic $pp$ collisions in the TDE debris. 
Furthermore, observed UV photons guarantee that efficient photomeson production occurs.   

Note that radio emission itself is attributed to synchrotron emission from electrons accelerated at the external forward shock. Inferred kinetic energy is $\sim10^{49}-{10}^{50}$~erg, but it can be larger if only a fraction of electrons are injected to acceleration. 
The external density is expected to be much lower than the wind density around this radius, and the amount of CRs accelerated during the observational time is expected to be small. However, CRs accelerated at the forward shock caused by the sub-relativistic outflow can be significant at later times~\citep[e.g.,][]{Zhang:2017hom}.

\subsubsection{Hidden jet model: unlikely}\label{sec:hidden-jet}
Neutrino emission from TDE jets has been considered by various authors~\citep[e.g.,][]{Murase:2008zzc,Wang:2011ip}. If the jet breaks out, bright X-ray emission is expected for an on-axis observer as seen in Swift\,J1644+57~\citep{2011Natur.476..421B,Bloom:2011xk}.
X-ray ``dark'' TDE jets were considered prior to the discovery of IceCube-191001A~\citep{Wang:2015mmh,Senno:2016bso}. In Figure~\ref{fig:nu} and Table~\ref{tab:nu}, we showed the prediction of the internal shock model of \cite{Senno:2016bso}. The model can account for at most ${\mathcal N}_{\nu_{\mu}}=10^{-3}\, (3\times10^{-4})$, under the optimistic assumption that the jet remains hidden for $3\times 10^{6}$~s. 

It has been speculated that a jet could be choked by TDE debris or wind~\citep{Wang:2015mmh}. The realization of such a setup is highly speculative, because the accreting material orbits around the SMBH and the jet may easily get launched almost perpendicular to the disk~\citep{Dai:2018jbr} without interacting with TDE debris. Nevertheless, it would be still useful to check whether hidden jets can be powerful or not. There are two relevant necessary conditions for having neutrino production in electromagnetically dark jets --- the radiation constraint and jet-stalling condition~\citep{Murase:2013ffa}. 
Shock acceleration is suppressed if the shock is radiation mediated. In the case of TDE jets, this can be satisfied. 
The second condition is more constraining, as explained below. 

As discussed in the previous section, it is possible to have a massive envelope consisting of the TDE debris, which can reprocess UV and X-ray radiation~\citep{Loeb:1997dv,DeColle:2012np}. 
For a density profile of $\varrho_{\rm in}{(r/R_{\rm in})}^{-3}$, \cite{Senno:2016bso} estimated the following upper limit on the isotropic-equivalent luminosity of a hidden jet,
\begin{equation}
L_j^{\rm iso}\lesssim 2\times10^{44}\,{\rm erg \, s^{-1}}~t_{\rm eng,6}^{-3}\varrho_{\rm in,-9.2}R_{\rm in,13.5}^{3}R_{\rm out,15.5}^2\theta_{j,-1}^{2},
\end{equation}
with half-opening angle $\theta_j$ powered by the central engine for duration of $t_{\rm eng}$. The jet is assumed to propagate through material spread between the inner radius $R_{\rm in}\sim R_T$ and outer radius $R_{\rm out}$.

Alternatively, optically-thick winds could help the jet get choked. Although the wind density in the jet (polar) direction would be lower than considered here, assuming that almost all the fallback material is expelled as a spherical outflow (see the previous section), the wind-driven debris density may be written as, $\varrho_{w}(r)=\dot{M}_{w}/(4\pi r^2V_w)\equiv D_wr^{-2}$, where $D_w=6.2\times{10}^{15}~{\rm g}~{\rm cm}^{-1}~\eta_{\rm fb}\eta_{\rm in,-1}^{-1/2} f_{T,-1.1}^{-5/12}M_{\rm BH,7}^{-5/6}M_{*}^{5/6+\xi}{(t/t_{\rm fb})}^{-5/3}$. 
With the wind outer radius $R_{\rm out}(\lesssim 0.5M_*V_w/\dot{M}_{w})$, the jet-stalling condition~\citep{Bromberg:2011fg,Mizuta:2013yma} gives the following upper limit on the isotropic-equivalent jet luminosity,
\begin{equation}
L_j^{\rm iso}\lesssim1.5\times10^{44}\,{\rm erg \, s^{-1}}~t_{\rm eng,6.5}^{-3}D_{w,15.8}R_{\rm out,16}^2\theta_{j,-1}^{2}.
\end{equation}
Thus, it is unlikely that powerful on-axis jets with $L_j^{\rm iso}\gtrsim{10}^{45}~{\rm erg}~{\rm s}^{-1}$ are X-ray dark. The hidden-jet model we considered satisfies the above constraints and predicts low neutrino fluxes, as shown in Figure~\ref{fig:nu} and Table~\ref{tab:nu}. 

Note that we do not exclude the possibility of off-axis jets. Optical polarimetry data of \tde are compatible with the existence of a spectral component which could be attributed to a jet, but cannot confirm such a structure~\citep{Lee_2020}. Also, neutrino emission from the off-axis jets is significantly lower.  
Jet-driven relativistic ejecta could also lead to quasi-isotropic neutrino emission but the luminosity will be largely diminished. 

Just recently, neutrino emission from \tde, in the presence of a hypothetical jet was discussed by~\citet{Winter:2020ptf}. Their assumed jet power violates the jet-stalling condition, so that the jet has to break out. Such a powerful on-axis jet is inconsistent with the observations. First, the sum of the properties of \tde set it apart from jetted TDEs such as Swift~J1644+57 and Swift~J2058+05 which had non-thermal X-ray spectra, $\sim 1000$ times brighter than the thermal spectrum of \tde. 
Second, once the jet breaks out, bright relativistic afterglow emission is also unavoidable~\citep[e.g.,][]{Generozov:2016oon}. According to the standard afterglow theory, with a conservative jet duration of $t_{\rm dur}\sim 3\times{10}^{6}$~s, the isotropic-equivalent energy becomes ${\mathcal E}_k\gtrsim 3\times{10}^{53}~{\rm erg}~L_{j,47}^{\rm iso}$.

\section{Contribution to the diffuse neutrino flux}
\label{sec:diffuse} 
TDE rates are expected to be $\sim{10}^{-5}-{10}^{-4}~{\rm gal}^{-1}~{\rm yr}^{-1}$, which correspond to ${10}^2-10^{3}~{\rm Gpc}^{-3}~{\rm yr}^{-1}$~\citep[e.g.,][]{Magorrian:1999vm,vanVelzen:2014dna,Sun2015ApJ...812...33S, Stone:2020vdg}. 
Noting that the CR energy per TDE carries only a fraction of the gravitational energy, $\mathcal{E}_{\rm CR}\lesssim 10^{50}-10^{52}$~erg, 
the diffuse neutrino flux is estimated to be
\begin{eqnarray}
E_\nu^2\Phi_\nu&\sim&1.7\times{10}^{-8}~{\rm GeV}~{\rm cm}^{-2}~{\rm s}^{-1}~{\rm sr}^{-1}~\left(\frac{2K}{1+K}\right)f_{\rm mes}\nonumber\\
&\times&\left(\frac{{\mathcal E}_{\rm CR,51}}{\mathcal{R}_{\rm CR}}\right)\left(\frac{\xi_z}{0.5}\right)
{\left(\frac{\rho_{\rm TDE}}{{10}^{2}~{\rm Gpc}^{-3}~{\rm yr}^{-1}}\right)}.\,\,\,\,\,\,\,\,\,\,\,
\label{eq:diffuse}
\end{eqnarray}
Here $K=1$ and $K=2$ for $p\gamma$ and $pp$ interactions, respectively, and $\xi_z\sim0.5$ is a factor representing the redshift evolution of TDEs~\citep[see][]{Waxman:1998yy,Murase:2016gly,Senno:2016bso}. 
Theoretically, TDEs are expected to have a negative redshift evolution~\citep{Sun2015ApJ...812...33S}, but it is highly uncertain observationally. If instead the TDE redshift evolution resembles the star-forming history that leads to $\xi_z\sim2-3$, TDEs can, in principle, make a significant contribution to the diffuse neutrino flux, allowing for uncertainties in the neutrino spectrum.

However, as mentioned in the Introduction, a stacking analysis of IceCube data found no counterparts to previously detected TDEs and concluded that they contribute at most $\sim30\%$ to the diffuse flux \citep{Stein:2019ivm}. Another important constraint comes from the non-detection of multiplet sources (i.e., the line-of-sight cumulative number of sources making signal multiplets). For the sources responsible for the diffuse flux, \citet{Senno:2016bso} gave
\begin{equation}
\label{eq:rho0}
\rho_0^{\rm eff}\gtrsim1.4\times10^{4}\,{\rm Gpc^{-3}}~{\rm yr^{-1}}\,\frac{{(b_mq_L/6.6)}^2{(T_{\rm IC}/6~{\rm yr})}^2}{{(\xi_z/0.5)}^3{\phi}_{\rm lim, -0.9}^3{(2\pi/\Delta \Omega)}^2}.
\end{equation}
where $\phi_{\rm lim}$ is the neutrino fluence limit,   $b_mq_L$ is a correction factor that depends on details of the analysis~\citep{Murase:2016gly}, and $\Delta\Omega$ is the field of view of the detector. 
Thus, although the results are rather sensitive to $\xi_z$, the above limit implies that not only jetted TDEs but also non-jetted TDEs are most likely to be sub-dominant in the diffuse neutrino sky. 

It is possible that a rare fraction of TDEs gives a sub-dominant contribution to the diffuse neutrino flux, but the above conclusion would hold when the neutrino emission is dominated by only a subset of all TDEs, as expected for radio-detected TDEs with relativistic or trans-relativistic outflows.
We argue that \tde could even be a TDE with off-axis jets (even though radio emission is dominated by the sub-relativistic wind component). The apparent rate density of jetted TDEs is $\rho_{\rm jetted}=0.03_{-0.02}^{+0.04}~{\rm Gpc}^{-3}~{\rm yr}^{-1}$~\citep{Sun2015ApJ...812...33S}, which leads to the true rate density, $R_{\rm jetted}\approx(2/\theta_j^2)\rho_{\rm jetted}\sim20~{\rm Gpc}^{-3}~{\rm yr}^{-1}~{(\rho_{\rm jetted}/0.1~{\rm Gpc}^{-3}~{\rm yr}^{-1})}\theta_{j,-1}^{-2}$. 
In this case, Equations~(\ref{eq:diffuse}) and (\ref{eq:rho0}) imply that TDEs with off-axis jets cannot dominantly account for the diffuse neutrino flux, but it is possible to give a contribution up to $\sim10$\%. Jetted TDEs could be even more common~\citep[e.g.,][]{DeColle:2012np,Coughlin:2013lva,Dai:2018jbr}, but the non-detection of several TDEs with radio observations implies that TDEs with relativistic outflows (jets or winds) are not ubiquitous. Note that this is a conclusion independent of the observation of \icnu.

\section{Summary and Discussion}\label{sec:summary}
We explored new possibilities of high-energy neutrino and gamma-ray emission from non-jetted regions in TDEs, focusing on the core regions (coronae and RIAFs) and outflows embedded in the TDE debris. 
We showed that in all considered models, efficient neutrino production via inelastic $pp$ collisions is possible, while $p\gamma$ interactions mainly on UV target photon fields are important for limiting the maximum CR energy and neutrino production at $\gtrsim100$~TeV energies. We also calculated CR-induced cascade electromagnetic emission, and found that $\gtrsim10$~GeV gamma-rays are attenuated in all models due to $\gamma\gamma\rightarrow e^-e^+$.
In the core models, we find that GeV gamma-rays are also suppressed, and the cascade emission can appear only at energies $\lesssim10$~MeV (100 MeV) for the corona (RIAF) model. However, in the hidden wind model, the cascade gamma-rays can still emerge in the GeV band.

We emphasize the importance of hard X-ray and soft gamma-ray observations to test the models considered in this work. In the corona model, only gamma-rays with $\lesssim10$ MeV can escape without attenuation due to the copious UV and X-ray photons in the disk and corona. In the RIAF model, the intrasource cascade emission emerges in the $\sim10-100$~MeV energy range. These gamma-rays could be further attenuated (and thereby re-emerge to even lower energies) if additional UV or X-ray photons exist outside the disk. 
Planned MeV satellites, such as \emph{e-ASTROGAM} ~\citep{DeAngelis:2016slk}, \emph{AMEGO}~\citep{Moiseev:2017mxg}, and \emph{GRAMS}~\citep{Aramaki:2019bpi} will be crucial for testing such models in the future. For example, \emph{AMEGO} with a sensitivity of $\sim10^{-12}~{\rm erg}~{\rm cm}^{-2}~{\rm s}^{-1}$ will be able to test the corona model for \tde-like objects at $\lesssim30-100$~Mpc.
Observations of non-jetted TDEs in hard X-rays during the optical/UV plateau phase (e.g., with late-time \emph{NuSTAR} observations) can give crucial constraints, although details depend on the attenuation in the surrounding debris (that is important especially in the soft X-ray band). The reasons are as follows. In the corona model, even if the X-ray luminosity is lower than the optical/UV luminosity, Comptonized emission by hot coronal electrons is generated in the hard X-ray band ($>10$~keV). In the RIAF model, around the critical mass accretion rate, the Comptonized emission by thermal electrons is expected in hard X-rays and soft gamma-rays~\citep{Kimura:2020thg}, and possible detection will enable us to measure the electron temperature. 
The hidden wind model may also predict hard X-ray photons through emission from the shock-heated material. The proton temperature reaches $kT_p\lesssim(3/16)m_pV_w^2$ in the shock downstream, and their energy should be transferred to electrons. If the balance between heating from protons and Compton cooling of electrons is achieved in the presence of copious photons, the immediate downstream may have an equilibrium temperature of $kT_e\sim 10~{\rm keV}~{(kT_p/100~{\rm keV})}^{2/5}~{(kT_{\rm OUV}/3~{\rm eV})}^{-8/5}n_{p,10}^{2/5}$~\citep{Murase:2010cu}. In the case of \tde, late-time upper limits on the 2-10~keV X-ray flux may already place stringent constraints on this model. Although details are beyond the scope of this work, dedicated searches for hard X-rays should be able to constrain possible dissipation caused by hidden jets or winds.  
Next-generation hard X-ray satellites, such as \emph{FORCE} \citep{2016SPIE.9905E..1OM}, will have a higher discovery potential. 

We stress that the purpose of this work is to provide a general framework of new models that are applicable to non-jetted high-energy emission from TDEs. 
We applied these models to \tde as an example, although their physical association with \icnu should be examined with great care. Keeping this in mind, we discussed the implications of \icnu that was detected $\sim 150$~days post-discovery of the TDE.
Whether this neutrino-TDE association is physical or not is still in question, but the reported significance is intriguing enough to make us discuss the implications of the neutrino-TDE connection. 
Overall, we conclude that explaining \icnu from \tde is typically challenging for all models in the literature, unless all TDEs emit neutrinos at the maximal rate consistent with our model predictions, in which the summed neutrino expectation could be $\sim0.1-1$. However, in principle, the corona model can be consistent with the data with $\lesssim 10 \% (1\%)$ detection probability of one neutrino. The required baryon loading is rather large, but not forbidding. This model predicts $\gtrsim100$~TeV neutrino emission preferencially from TDEs with heavy SMBHs, and the neutrino detection around when the plateau emission was found could be associated with the time of formation of the disk-corona structure. Intriguingly, the host galaxy of AT2019dsg is expected to have such a heavy SMBH. 
In the RIAF model, the detection probability is less than 1\%, so it is not easy to accommodate the neutrino detection. 
It is also challenging for the hidden wind model to satisfactorily explain the neutrino detection, especially when the bolometric correction on the CR luminosity is taken into account, but the model is not ruled out by the observations. An advantage of this model is that radio observations showed the existence of a powerful sub-relativistic outflow. Interestingly, the CR-induced cascade gamma-ray spectrum has a peak in the GeV range. The predicted flux is still below the \emph{Fermi}-LAT upper limit for \tde, but our results imply that cascade gamma-rays provide one of the promising tests for the hidden wind model. 
Finally, we discussed the hidden jet scenario, where electromagnetic emission from the jet is hidden by TDE debris. In this model, the jet-stalling condition provides an upper limit on the jet power, in which the resulting neutrino flux should be as low as in our RIAF model. More luminous jets would break out from the obscuring material, which become inconsistent with the absence of bright X-ray and afterglow emission. Hence, this scenario is unlikely to account for the multi-messenger observations of \tde. As in the case of the jetted AGN, TXS 0506+056~\citep{Keivani:2018rnh}, we still lack a convincing picture to explain the multimessenger and multiwavelength data. 

When the system is nearly calorimetric (as typically expected in corona, hidden wind, and hidden jet models), CRs are depleted without contributing to the observed CR flux. However, the external shock formed by sub-relativistic outflows at large radii can produce very high-energy CRs that could contribute to the observed CR flux~\citep{Zhang:2017hom}. Alternatively, the external shock formed by TDE jets or internal shocks of possible low-luminosity TDEs may accelerate nuclei up to ultrahigh energies~\citep{Zhang:2017hom,Guepin:2017abw}. In the RIAF model, CRs may leave the disk without significant energy losses~\citep{Kimura:2014jba}. Escaping CRs further interact with circumnuclear material, making neutrinos and gamma-rays~\citep{Fujita:2015xva}. 

Physical phenomena in TDEs are not fully understood, and various processes that occur at $\lesssim10^{15}-10^{16}$~cm are highly uncertain. Open issues include the details of circularization, the roles of stream crossing shocks, disk formation, outflow dynamics, and the properties of disk-driven winds. One of the difficulties comes from the fact that soft X-rays can be obscured by TDE debris and may be reprocessed to optical/UV emission there. High-energy neutrinos and gamma-rays can be used as a unique probe of dissipation in the complex regions embedded in the TDE debris, and our work demonstrated that neutrino and gamma-ray observations can shed new light on these mysterious phenomena especially through future high-energy neutrino observations with IceCube-Gen2~\citep{Aartsen:2014njl}, KM3Net~\citep{Adrian-Martinez:2016fdl}, and electromagnetic observations with planned hard X-ray and MeV gamma-ray satellites.

{\it Note added:}
While this paper was being prepared, the work of \citet{Winter:2020ptf} came out. 
Our main non-jetted models are different from theirs, and neutrino production in this paper does not rely on the uncertain X-ray emission of the \tde, in contrast to their work that uses X-ray photons as the main targets.
For the hidden jet model, this work relies on earlier predictions of \cite{Senno:2016bso}, according to which the jet-stalling condition forbids a powerful on-axis jet, like the one postulated by \citet{Winter:2020ptf}.


\begin{acknowledgements}
K.M. thanks Kimitake Hayasaki and Ryo Yamazaki for useful discussion. 
We also thank an anonymous referee for helpful suggestions. 
The work is supported by the Alfred P. Sloan Foundation, NSF Grant No.~AST-1908689, and JSPS KAKENHI 
No.~20H01901 (K.M.), and JSPS Research Fellowship and JSPS KAKENHI No.~19J00198 (S.S.K.).
B.T.Z. is supported by the IGC Fellowship at Penn State, and M.P. acknowledges support from the Lyman Jr.~Spitzer Postdoctoral Fellowship and NASA Fermi grant No.~80NSSC18K1745.
\end{acknowledgements}

\bibliographystyle{aasjournal}
\bibliography{kmurase,tde}

\end{document}